\documentclass[conference,hidelinks]{IEEEtran}
\IEEEoverridecommandlockouts

\usepackage{amsmath,amssymb,amsfonts}
\usepackage{algorithmic}
\usepackage{graphicx}
\usepackage{textcomp}
\usepackage{xcolor}
\usepackage{booktabs}
\def\BibTeX{{\rm B\kern-.05em{\sc i\kern-.025em b}\kern-.08em
    T\kern-.1667em\lower.7ex\hbox{E}\kern-.125emX}}

\usepackage[style=ieee,mincitenames=1,maxcitenames=2,minbibnames=1,maxbibnames=3]{biblatex}
\AtBeginBibliography{\footnotesize}
\AtEveryBibitem{\clearlist{location}}

\usepackage{hyperref}
\usepackage{cleveref} 
\usepackage{glossaries}
\usepackage{booktabs, multirow, multicol}
\usepackage{graphicx}
\usepackage{tabularx}
\glsdisablehyper

\usepackage{orcidlink}

\newcommand*{\CERTH}{
\textit{ITI-CERTH}, Thessaloniki, Greece
}
\newcommand*{\IDMT}{
\textit{Fraunhofer IDMT}, Ilmenau, Germany
}

\makeatletter
\newcommand{\newlineauthors}{%
  \end{@IEEEauthorhalign}\hfill\mbox{}\par
  \mbox{}\hfill\begin{@IEEEauthorhalign}
}
\makeatother

\newacronym{ai}{AI}{Artificial Intelligence}
\newacronym{cpra}{CPRA}{California Privacy Rights Act}
\newacronym{dsp}{DSP}{Digital Signal Processing}
\newacronym{fl}{FL}{Federated Learning}
\newacronym{gdpr}{GDPR}{General Data Protection Regulation}
\newacronym{mos}{MOS}{Mean Opinion Score}
\newacronym{nonIID}{non-IID}{non-independent and identically distributed data}
\newacronym{xai}{XAI}{Explainable Artificial Intelligence}
\newacronym{mfcc}{MFCC}{Mel Frequency Cepstral Coefficients}
\newacronym{tts}{TTS}{Text-To-Speech}
\newacronym{vc}{VC}{Voice Conversion}
\newacronym{cnns}{CNNs}{Convolutional Neural Networks}
\newacronym{stft}{STFT}{Short-time Fourier Transform}
\newacronym{dnn}{DNN}{Deep Neural Network}
\newacronym{la}{LA}{Logical Access}
\newacronym[plural=GMMs,longplural={Gaussian Mixture Models}]{gmm}{GMM}{Gaussian Mixture Model}

\makeatletter
\newcommand{\warn}[1]{\@latex@warning{#1}}
\makeatother

\begin{document}

\title{Open Challenges in Synthetic Speech Detection\\
\thanks{This paper was supported by the EU H2020 AI4Media project (grant no 951911)%
\textcolor{black}{ ~and by the BMBF SpeechTrust+ project (grant no 13N16267)}%
.}
}

\author{
\IEEEauthorblockN{Christoforos Papastergiopoulos\orcidlink{0000-0002-4130-4866}}
\IEEEauthorblockA{\CERTH\\papasterc@iti.gr}
\and
\IEEEauthorblockN{Artem Yaroshchuk}
\IEEEauthorblockA{\IDMT\\artem.yaroshchuk@idmt.fraunhofer.de}
\and
\IEEEauthorblockN{Anastasios Vafeiadis\orcidlink{0000-0003-1102-5708}}
\IEEEauthorblockA{\CERTH\\anasvaf@iti.gr}
\newlineauthors
\IEEEauthorblockN{Luca Cuccovillo\orcidlink{0000-0001-5559-6508}}
\IEEEauthorblockA{\IDMT\\luca.cuccovillo@idmt.fraunhofer.de}
\and
\IEEEauthorblockN{Patrick Aichroth\orcidlink{0000-0003-4777-6335}}
\IEEEauthorblockA{\IDMT\\patrick.aichroth@idmt.fraunhofer.de}
\and
\IEEEauthorblockN{Konstantinos Votis}
\IEEEauthorblockA{\CERTH\\kvotis@iti.gr}
\newlineauthors
\IEEEauthorblockN{Dimitrios Tzovaras}
\IEEEauthorblockA{\CERTH\\dimitrios.tzovaras@iti.gr}
}

\author{
\IEEEauthorblockN{%
  Luca Cuccovillo\IEEEauthorrefmark{1}\orcidlink{0000-0001-5559-6508},
  Christoforos Papastergiopoulos\IEEEauthorrefmark{2}\orcidlink{0000-0002-4130-4866},
  Anastasios Vafeiadis\IEEEauthorrefmark{2}\orcidlink{0000-0003-1102-5708}, 
  Artem Yaroshchuk\IEEEauthorrefmark{1},\\
  Patrick Aichroth\IEEEauthorrefmark{1}\orcidlink{0000-0003-4777-6335},
  Konstantinos Votis\IEEEauthorrefmark{2}, and
  Dimitrios Tzovaras\IEEEauthorrefmark{2}
}
\IEEEauthorblockA{}
\IEEEauthorblockA{\IEEEauthorrefmark{1}\textit{Fraunhofer Institute for Digital Media Technology}, Ilmenau, Germany}
\IEEEauthorblockA{\IEEEauthorrefmark{2}\textit{Centre for Research and Technology Hellas}, Thessaloniki, Greece}
}
\maketitle

\begin{abstract}
In this paper the current status and open challenges of synthetic speech detection are addressed.
The work comprises 
an initial analysis of available open datasets and of existing detection methods, 
a description of the requirements for new research datasets compliant with regulations and better representing real-case scenarios, and
a discussion of the desired characteristics of future trustworthy detection methods in terms of both functional and non-functional requirements.
Compared to other works, based on specific detection solutions or presenting single dataset of synthetic speeches, 
our paper is meant to orient future state-of-the-art research in the domain, to quickly lessen the current gap between synthesis and detection approaches.


\end{abstract}


\section{Introduction}
With the advent of \gls{tts} and \gls{vc} methods based on \gls{ai}, it is now possible to generate synthetic speech of such a high quality that is indistinguishable from real voices by human listeners \cite{vocoders_hifigan2020,vocoders_waveglow2019,vocoders_wavenet2016,tts_shen2018natural}. 

This new level of quality is enabling the birth of ground-breaking applications, such as allowing speech-impaired people who lost their voices during their adulthood to recover them using past audio recordings~\cite{parrotron}. The \emph{misuse} of speech synthesis to impersonate another person, however, brings several dangers, among which the possibility of gaining illicit access to private data from close relatives, or the chance of discrediting private citizens or public figures, e.g., by letting their pronounce racist or homophobic statements.

The danger of such misuses highlights the need for \gls{ai} solutions that can discern synthetic audio from real audio, especially when the individual freedoms are at stake in court cases. The creation of such \gls{ai} solutions would bring great benefits to the audio forensics community and to the media verification one. 

Therefore, the research community reacted by proposing creative and innovative detection methods~\cite{alegre2013one,ma2021rw,chettri2019ensemble,borrelli2021synthetic,albadawy2019detecting,papasterc_2022_icmr,conti2022emotions}, as-well-as by releasing open datasets for training and evaluation of present and past technologies~\cite{datasets_wavefake2021,datasets_asvspoof2021,datasets_halftruth2021,datasets_fakeorreal2019,datasets_add2022}. 

The current state of the art in the speech synthesis detection domain, however, suffers from several issues. The available open datasets, despite the remarkable efforts required by their collection, still lack real and synthetic data meeting the technical requirements for the production of a robust detection system that can be used in the industry. There is a scarce collaboration between industrial and research institutions, pushing researchers to train and validate their detection methods on data which does not match the requirements of the final users their researches are meant for. Model outputs often lack interpretability, which is a key requirement in many court cases for the analysis to be admittable.

In this paper we thus propose to focus on collecting requirements for new research datasets compliant with regulations and better
representing real-case scenarios, on defining characteristics necessary for trustworthy and interpretable detection methods in terms
of both functional and non-functional requirements, and on the synergies that arise from an appropriate requirement analysis of future actions for dataset creation and synthesis detection. 

The rest of the paper is organized as follows.
In \Cref{sec:open-datasets} we analyze the available open datasets of synthetic speech, highlighting the respective strengths and potential issues of the content thereof. 
In \Cref{sec:detection-algorithms} we then focus on selected state-of-the-art detection algorithms, relating them to the aforementioned datasets and describing their most characteristic features. 
In \Cref{sec:open-challenges}, the core of this work, we leverage the previous information to outline existing challenges to be faced by future state-of-the-art research in the domain, to quickly lessen the current gap between synthesis and detection approaches. 
\Cref{sec:conclusions} closes with a summary of the paper and conclusions driven by the whole work.
\section{Open Datasets of Synthetic Speech}\label{sec:open-datasets}
Various research institutions and individuals have contributed in the synthetic speech detection field by creating open datasets for training and benchmarking. This section describes the most popular ones, summarised at the end in \Cref{tab:datasets_overview}.

\subsection{The ASVspoof DF Challenge Dataset}
The international challenge on Automatic Speaker Verification Spoofing detection (ASVspoof) aims at promoting the study of spoofing attacks against automatic speaker verification systems and their countermeasures. The datasets used in ASVspoof contain both bona fide speech and synthetic speech utterances, generated using a number of different \gls{tts} and \gls{vc} systems~\cite{datasets_asvspoof2021}. The first ASVspoof challenge took place in 2015, and as the research grew on the field of \gls{tts} synthesis, the ASVspoof databases were revised to increase the difficulty of the tasks. In its 2021 version, the ASVspoof challenge comprises three tasks, namely the \gls{la} task focusing on the detection of bona fide and spoofed utterances in communication networks, the Physical Access (PA) task focusing on the detection of replay attacks, and the speech DeepFake (DF) task, which focuses specifically on the detection of synthetic speech, without any relation to the speaker recognition use case. 

The ASVspoof DF dataset is based upon the Voice Cloning Toolkit (VCTK) corpus~\cite{yamagishi2019cstr} and includes synthetic speech and converted voice signals generated with a total of 17 state-of-the-art algorithms, among which 6 are known and 11 are kept undisclosed. The DF database, which focuses on the problem of synthetic speech detection, is divided into two parts: the attacker side, and the defender side. 

For the defender side, the training data contains 2,580 bona fide utterances and 22,800 spoofed utterances generated by 4 \gls{tts} algorithms. The development partition contains 1,484 bona fide target utterances, 1,064 bona fide non-target utterances, and 22,296 spoofed utterances generated with the same \gls{tts} and \gls{vc} algorithms. The setup for the evaluation set is similar to that for the development set, containing 5,370 bona-fide target utterances and 1,985 bona fide non-target utterances. Spoofed data comprises 63,882 utterances generated by using 7 \gls{tts} and 6 \gls{vc} spoofing algorithms. The training, development, and evaluation sets include 20 training speakers, 10 target and 10 non-target speakers, and 48 target and 19 non-target speakers, respectively \cite{datasets_asvspoof2021}. 

The attacker side contains unseen training data in order to build \gls{tts} and \gls{vc} systems that can generate speech with similar characteristics to those of the target speakers, with the idea that every participant may augment its training set by introducing data generated by additional systems. For every target speaker, there are 200 utterances that are unique and not included in defender side subsets.

ASVspoof is at present the corpus closest to satisfying the requirements of a ``correct'' dataset for training and evaluation of synthesis detection, due to its large variety of speakers, and to the presence of several synthesis algorithms. Its main drawback is that, given the challenge-driven nature of the dataset, the synthesis algorithms involved remain undisclosed. This choice, perfectly in-line with the philosophy of a challenge with competitors, is however detrimental for a systematic development of countermeasures: The interactions between feature extraction networks, voice conversion techniques and related vocoders cannot be studied explicitly, despite being possibly already reflected in the distribution of the data.

\subsection{The Fake-or-Real (FoR) Dataset} 
The Fake-or-Real (FoR) dataset contains more than 117,000 real speech utterances collected from several open source datasets and 87,000 synthesized phrases from the English-French translations dataset generated using commercial and open source TTS systems \cite{datasets_fakeorreal2019}. It is available in four formats, differentiated according to the respective pre-processing methodology. To avoid overfitting to specific voices or recording devices the dataset is gender-balanced and incorporates real utterances from Arctic~\cite{datasets_arctic2004}, VoxForge~\cite{voxforge}, LJ Speech datasets. The synthesized voices, however, are not present in the real subset, which leaves the possibility for a machine learning algorithm to classify based on these specific voices, instead of learning the differences between generated and real utterances.

\subsection{The WaveFake Dataset}
WaveFake provides a collection of generated audio clips using several flow-based and Generative Adversarial Network architectures~\cite{datasets_wavefake2021}. The dataset is sampled by extracting Mel spectrograms from LJ Speech~\cite{datasets_ljspeech2017} and JSUT~\cite{datasets_jsut2017} datasets and feeding them to the vocoder models trained on the same distributions. The TTS samples are generated using Mozilla Common Voice Corpus. Both LJ Speech and JSUT, however,  comprise a single female speaker, which implies that the generated distribution does not represent a variety of voices and recording conditions.

\subsection{The ADD Challenge Dataset}
The first Audio Deep Synthesis Detection (ADD) Challenge~\cite{datasets_add2022} includes three tasks: the low-quality fake audio detection (LF) task; the partially fake audio detection (PF) task; the audio fake game (FG) task. ADD 2022 addresses several attack scenarios not covered by ASVspoof, e.g., fake utterances containing disturbances and background noises. The ADD challenge dataset is based on AISHELL-3~\cite{datasets_aishell2021} -- a multi-speaker Mandarin speech corpus and provides speaker-disjoint training, development and adaptation subsets composed of real and fake utterances containing various noises. While a base dataset incorporates a large variety of voices, it is unbalanced with respect to gender, accent and age group and recorded using the same equipment setup.
Unfortunately, the dataset cannot be redistributed in original nor in any derivative forms, but only downloaded for non-commercial purposes by contacting directly the creators. The specific synthesis methods used during the generation have not been revealed by authors at present, and therefore the dataset presents the same weakness of the ASVspoof DF one: Low usability for systematic research.

\subsection{The Half-Truth Dataset}
The Half-Truth dataset~\cite{datasets_halftruth2021} is designed to promote research on detecting audio utterances which are only partially fake. Similarly to the ADD challenge dataset it is based on AISHELL-3 corpus, and includes not only fully-synthetic utterances, but also partially fake ones, obtained by manipulating the original speech by splicing segments of synthetic audio. The fake audio, however, is generated using a \emph{single} text-to-spectrogram and vocoder pipeline. Therefore, the dataset targets a problem more difficult than distinguishing completely synthesized and completely bona fide utterances, but unfortunately lacks coverage of a wide variety of synthesis methods.


\begin{table*}[ht]
\centering
\caption{Available datasets for the development of spoofing attacks detection models.}\label{tab:datasets_overview}
\begin{tabularx}{0.95\textwidth}{@{}lllX@{}}
\toprule
Dataset       & Real Utterances & Fake Utterances & Notes \\ \midrule
FoR~\cite{datasets_fakeorreal2019}          & 117,000         & 87,000          & Gender balanced, class balanced and truncated versions  of the original dataset are provided.                             \\
ASVSpoof~\cite{datasets_asvspoof2021} & 5,128            & 25,096           & Different datasets are provided to tackle three major forms of spoofing  attacks, namely replay, voice conversion and speech synthesis. \\
WaveFake~\cite{datasets_wavefake2021}      & -               & 117,985         & Fake utterances include a single female voice, resulting in data distribution bias.                                                  \\
ADD~\cite{datasets_add2022}           & 5,319            & 45,367           & Low usability for systematic research due to restrictions in distribution.                                                                                                        \\
Half-Truth    & 26,554           & 26,554           & Partially fake utterances are provided  for the purpose of detecting  manipulated real audio.\\ \bottomrule
\end{tabularx}
\end{table*}
\section{Synthetic Speech Detection Algorithms}\label{sec:detection-algorithms}

Detecting whether a speech recording belongs to a human or is synthetically generated is considered a challenging task. There is a wide variety of different methodologies to synthesize human speech and each one has its own unique characteristics \cite{tts_shen2018natural}. The speech synthesis quality is measured by the \gls{mos} and most methods use a common benchmark dataset for evaluation.

In order to identify a synthetically generated speech recording, a number of different algorithms has been proposed. These include traditional machine learning classifiers that use hand-crafted audio features as input \cite{alegre2013one} and deep neural networks that can provide an end-to-end solution based on raw speech recordings as input \cite{ma2021rw}.

The most common evaluation metrics to measure the performance of the algorithms are Equal error rate (EER) and tandem detection cost function (t-DCF). Achieving a lower EER translates into better performance in spoofing detection. On the other hand, t-DCF measures the influence of spoofing detection on the reliability of an Automatic Speak Verification (ASV) system, with a lower t-DCF score implying that the ASV system is more reliable~\cite{datasets_asvspoof2021}.


\subsection{Hand-Crafted Features with Traditional Classifiers}
Meaningful acoustic features from speech samples are important to discriminate between real and synthetic audio clips. It has been shown that hand-crafted features tailored for the task of voice spoofing detection could outperform, on specific datasets, more complex classifiers \cite{chettri2019ensemble}. For instance, \citeauthor{borrelli2021synthetic}~\cite{borrelli2021synthetic} showed that combining short-term and long-term predictors outperformed the baseline method \cite{albadawy2019detecting} on the ASVspoof 2019 dataset.

The most popular acoustic features that are commonly used on datasets that contain human speech are the \gls{mfcc}. The MFCCs are based on the mel scale, which resembles the way which the human ear perceives sound. Other popular examples are the constant-Q cepstral coefficients, extracted from the power spectrum of the windowed speech frames, the linear frequency cepstral coefficients and linear predictive coding coefficients have also been explored for the task of spoofed speech detection. These features are usually used as input to supervised classifiers, such as \glspl{gmm} \cite{bakar2018experimental} and support vector machines \cite{arif2021voice}.

\subsection{Deep Learning in Synthetic Speech Detection}
With the recent advances in affordable processing units, deep learning has received significant research interest in the field of spoofing attack detection. The ability of deep neural networks to extract meaningful features from the raw input has resulted in building end-to-end systems capable of recognizing synthetic audio~\cite{dinkel2017end} and replay attacks~\cite{tom2018end}.

Popular methods are based on applying state-of-the art \gls{cnns} from the computer vision area for the task of synthetic voice detection. In \cite{wang2020deepsonar}, the authors have used a thin-ResNet architecture and classified whether a recording is fake or real by monitoring the layer-wise neuron behaviors. The so-called ``DeepSonar'' architecture was applied in three datasets of two different languages (English and Chinese), specifically the publicly available FoR dataset, a self-generated English dataset based on the \gls{vc} Sprocket toolkit~\cite{kobayashi2018sprocket} and a self-generated Chinese dataset based on the Baidu speech synthesis system~\cite{peng2020non}, and showed promising generalization capabilities. \citeauthor{zhang2021fake}~\cite{zhang2021fake} introduced a fake speech detection architecture based on a ResNet combined with a transformer encoder. They applied five popular audio data augmentation methods, namely  Gaussian noise addition, signal-to-noise ratio noise
addition, time shifting, pitch shifting, and time stretching, and extracted spectral features based on magnitude \gls{stft} representations to train the transformer encoder ResNet. The network was evaluated on the FoR and the ASVspoof 2019 LA datasets separately and also cross-dataset in order to test the generalizability of the proposed method. In \cite{papasterc_2022_icmr} the authors have used linear STFT, mel-spectrogram, and MFCC magnitude representations and a combination of the above three as a stacked input to train a VGG16 network on the FoR dataset and evaluate the pre-trained architecture on a synthetic dataset based on the TIMIT corpus~\cite{garofolo1993timit}. Statistical analysis was performed on the FoR dataset to show the discrepancy between the training/validation and the testing subsets. \citeauthor{tak2021end}~\cite{tak2021end} used raw waveforms as input to a network architecture consisting of sinc filters, residual blocks and a recurrent neural network based on gated recurrent units, resulting in an end-to-end approach for synthetic speech detection. Results showed that the so-called ``RawNet2'' classifier can achieve competitive performance in a subset of the ASVspoof 2019 dataset. In a later work, the authors improved upon RawNet2 by using a spectro-temporal graph attention network (GAT)~\cite{tak2022gat} which is able to learn the relationship between cues spanning across different sub-bands and temporal intervals. The proposed RawGAT-ST model achieved an EER of $1.06$\% in the ASVspoof 2019 logical access database.  Most recently, \citeauthor{conti2022emotions}~\cite{conti2022emotions} presented a novel method for synthetic speech detection based on high-level semantic feature extraction. Their focus was on detecting deepfake speech tracks generated with \gls{tts} algorithms exploiting their emotional voice content. The developed system was evaluated on five public datasets, namely the ASVspoof 2019, LibriSpeech~\cite{panayotov2015librispeech}, LJ Speech~\cite{datasets_ljspeech2017}, Cloud 2019~\cite{lieto2019hello} and the speech tracks of the Interactive Emotional Dyadic Motion Capture~\cite{busso2008iemocap}.

Most of the aforementioned approaches try to reach high performances on public datasets, while being computationally expensive. An attempt to use the concept of transfer learning to create an efficient CNN, using a Z-normalized log magnitude STFT spectrogram as input was done in~\cite{subramani2020learning}. The proposed models needed fewer than 50,000 parameters and had approximately 100 KB of memory footprint.



\section{Open Challenges and Research Questions}\label{sec:open-challenges}
\subsection{Collection of Real Audio Data}
The first challenge related to the collection of data suitable for research on synthetic speech detection, is the collection of \emph{real} data. 

The distribution of the real data should be a good sample of the general target population. That means that is should be balanced in terms of gender and age, that each speaker should be present with an equal amount of content, and ideally that all dialects of the target language should be equally represented. In terms of quantity, each voice should be represented with enough transcribed examples to train -- or at least fine-tune -- the \gls{tts} systems and vocoders they are meant to be related with. This kind of efforts goes beyond the straightforward reuse of datasets developed for speech-to-text applications, which lack balance and in some cases -- such as the ubiquitous LJ Speech dataset -- refer to a single speaker. 

Furthermore, in order for a dataset to be used by researcher of every country, it is necessary that the collection process is performed adhering to legal constraints for data protection, such as the European \gls{gdpr}, the upcoming US-state \gls{cpra}, and future initiatives in a similar direction. It is necessary to consider that the presence of data publicly available on social networks or file sharing platforms does not give researchers the rights to \emph{use} the content without consent of the people depicted or recorded: Scavenging files from the internet, even if done with the good intention of countering the abuse of synthetic speech, leads to datasets which are illegal to use by large parts of the research community. 

Lastly, real data should take into account the impact of the recording equipment: To avoid incurring into domain-adaptation issues \cite{daume2006domain} whenever deploying detection algorithms in real-life, the recordings in the dataset should be acquired in dry controlled environments and with high quality equipment, so that more challenging conditions can be created by proper signal processing chains for data augmentation. Using consumer recording devices in reverberant environments might introduce unexpected features in the data distribution, which eventually may lead to systems identifying these hidden regularities, rather than traces left by the synthesis process.

\subsection{Creation of Synthetic Audio Data}
A second challenge related to the collection of data suitable for research on synthetic speech detection, is the creation of representative \emph{synthetic} data.

Synthetic data has to be paired with the real data it should be compared with: All synthetic voices should have a natural equivalent, so that the same balance expected from real data is also found in their synthetic counterparts.

It is also important, however, to consider the impact and variability of audio synthesis components. The vocoders involved, for instance, react differently to the data according to how they have been trained or fine-tuned. Whereas vocoders trained on a single voice sometimes introduce clear artifacts if applied to different voices, vocoders trained on multi-speaker datasets tend to reach a higher quality during the synthesis of voices outside the training set. The highest quality can be reached when the vocoders have been trained or fine-tuned for the sole voice they are meant to synthetize: Data produced with ad-hoc vocoders, albeit costly due to the expensiveness of the training phase, is also the most interesting in terms of detection, and poses a real challenge in terms of research. 

Similar considerations should be made about components for \gls{tts} feature extraction or voice conversion. Data produced by these models should reflect usage in naive conditions in which singles-speakers pre-trained models are applied directly, as well as usage after fine-tuning on both low quality and high quality variants of the target speaker to be impersonated. Even more, these variants should all be chained with all variants of the vocoders for non-end-to-end systems, leading to a complexity which is challenging to achieve, but would well represent the data distribution of the algorithms for speech synthesis which are available to the public -- and are therefore candidates for becoming vectors of impersonation attacks.

A last requirement to assure a relevant data distribution of synthetic examples, especially for what concerns \gls{tts} feature extraction components, is to maximize the expressiveness achievable by these engines. Since the very early experiments with WaveNet, the authors of these models noticed that even though their networks could synthesize speech samples with natural segmental quality, the samples sometimes had unnatural prosody by stressing wrong words in a sentence, especially in the case of very short sentences of a few words: The size of the receptive field of the WaveNet, 240 milliseconds, was not long enough to capture long-term dependency of F\textsubscript{0} contours~\cite{vocoders_wavenet2016}. To achieve the highest possible naturalness of the synthetic audios, 
synthetic data should be created using long periods, of duration higher than a few seconds, to avoid wrong prosodies and to minimize their lack of expressiveness.

\subsection{Algorithms Generalization}
As spoofing attacks can be unpredictable in nature, it is important to develop robust models able to identify the relevant artifacts in speech recordings and generalize well to unknown scenarios. \citeauthor{balamurali2019toward}~\cite{balamurali2019toward} highlighted the importance of combining a variety of traditional audio features with a \gls{gmm} classifier in order to generalize in replay attacks, on the ASVspoof 2017 dataset. Analyzing the results of relevant challenges in the AI research community, such as the ASVspoof challenge, it can be observed that participants prioritize achieving state-of-the-art performance in the development set that is provided, by optimizing their models through overfitting to the dataset at hand. The benchmark scores of the models strengthen this hypothesis, since all of them showcase a decrease in performance when tested on novel data with divergent audio characteristics and different pre-processing methods~\cite{muller2022does,chen2020generalization}.

In order to improve the generalization ability of synthetic speech detection models, the research efforts should be extended from audio feature extraction methodologies to the investigation of novel \gls{dnn} architectures. Moreover, new audio features, which do not focus on the specific vocal characteristics (e.g., pitch and timbre) of the respective training dataset, can guide models to learn more generic information to detect synthetic speech.

\subsection{Federated Learning}


In recent years, data regulations and privacy concerns for data sharing among different parties of a network, increased the demand for distributed learning methodologies. \gls{fl} enables the collaborative training of a neural network from siloed data centers and remote devices, while preserving data privacy by avoiding the exchange of data samples. Although \gls{fl} resolves the main issue of maintaining data privacy, new challenges arise by the heterogeneity of the data distribution among the network nodes. This results in training a \gls{fl} model with \gls{nonIID}, resulting in a reduced learning effectiveness compared to centralised training methods~\cite{li2021federated}. There are methods in literature trying to address the issue~\cite{zhao2018federated, guliani2021training}, but the problem is still an open research question. 

\gls{fl} can speed up the development of a robust collaborative model able to detect fake speech, by eliminating the time spent by legal departments to ensure compliance with data regulations when sharing audio recordings of real people. For the model to meet industrial standards when it comes to its performance, more effort should be put in the research of solutions for the \gls{fl} \gls{nonIID} issue in the audio domain. Thus, the need for an open source dataset that can allow researchers to simulate a distributed learning environment where both synthetic speech samples and real audio samples have different data distributions for each participant. This requires the utilization of a variety of different \gls{tts} algorithms as well as the collection of real audio samples that are phonetically rich, gender balanced and cover a wide range of different dialects. 

\subsection{Explainability}

The advancements of \glspl{dnn} and their widespread adoption in many fields of high relevance for the industry -- such as computer vision and natural language processing -- increased the social impact of AI models and led to a need for regulatory frameworks such as the ``right to explanation'' of AI systems promoted by the European Union~\cite{goodman2017european}:
In order for legal bodies to oversee and regulate the proper use of AI technologies, the necessity for methods that can explain the inner workings of black box AI models emerged. 

Many AI-based proposals in the audio domain builds upon the breakthroughs of AI research in the computer vision domain. Consequently, \gls{xai} methods that focus on the interpretability of image classification tasks can be utilized for the explainability of speech synthesis detection models. Several gradient-based methods in literature (e.g., Grad-CAM~\cite{selvaraju2017grad}) provide visual explanations of the decision making process of \gls{cnns}. However, when the inputs of the network are image representations of audio data (e.g., STFT and mel spectrograms), the visual explanations that these \gls{xai} algorithms provide are ambiguous and prone to subjective assessment. 

As follows, novel \gls{xai} algorithms that are explicitly designed for audio classification tasks can provide insights and assist in the improvement of the existing AI solutions. For instance, in the synthetic audio detection problem, understanding the distinct voice characteristics that distinguish synthetic speech samples from real audio samples can be leveraged to improve the generalizability of the model. Moreover, it can guide the data processing and collection procedures in the development of a new fake or real dataset, with the aim of being feature rich and adequately unbiased, and hence allowing the training of better classification models.



\section{Conclusions}\label{sec:conclusions}

To our knowledge, this work is the first one trying to collect requirements for synthetic speech detection, in terms of both technical requirements of the related development and evaluation datasets, and in terms of functional and non-functional requirements of future trustworthy and interpretable analysis methods. 

The analysis of the current state of the art in the speech synthesis detection domain allowed us to streamline the existing challenges to be faced by researchers in this field. The goal we pursued was to set a solid basis for future works, useful to lessen the gap between synthesis and detection approaches. In addition to detailing the requirements of the real and synthetic data to be collected and released -- including legal regulations which if ignored may undermine many potential collaborations -- we also highlighted the need for interpretability of the detection system to be developed, and showed how federated learning might be use as collaborative technique to quickly counter the constant flow of new synthesis methods released to the public, if problems related to the presence of \gls{nonIID} training data distributions are considered in advance.

In the near future we are going to start releasing a dataset of natural and synthetic recordings, taking into account the requirements we identified within these pages and adhering to the international legal regulations. Such an open dataset could be used for development or benchmarking altogether, and to encourage collaborations between and within research and industrial institutions.

\warn{TODO: Balance last page for camera-ready version}
\printbibliography

\end{document}